\documentclass[12pt]{article}

\usepackage[a4paper,margin=3cm]{geometry}
\usepackage[T1]{fontenc}

\usepackage{authblk}

\usepackage{sectsty}
\sectionfont{\fontsize{16}{20}\selectfont}
\subsectionfont{\fontsize{14}{18}\normalfont\em\selectfont}

\usepackage{titlesec}
\titlelabel{\thetitle.\,}

\usepackage[parfill]{parskip}
\usepackage{setspace}

\usepackage[labelfont=bf,font={small,stretch=0.95}]{caption}

\usepackage[super,sort,compress,numbers]{natbib}
\usepackage{hyperref}

\usepackage{gensymb}
\usepackage{amsmath}
\usepackage{amssymb}

\usepackage{graphicx}
\usepackage{multirow}

\usepackage[version=4]{mhchem}

\usepackage{mathptmx}

\begin{document}

\title{\fontsize{20}{26}\bf\selectfont Machine-learning-driven modelling of amorphous and  polycrystalline \ce{BaZrS3}}

\author[1]{Laura-Bianca~Pa\c{s}ca}
\author[1]{Yuanbin~Liu}
\author[1,2]{Andy~S.~Anker}
\author[1]{Ludmilla~Steier}
\author[1]{Volker~L.~Deringer\thanks{volker.deringer@chem.ox.ac.uk}}

\affil[1]{Inorganic Chemistry Laboratory, Department of Chemistry, University of Oxford, \protect\\ Oxford OX1 3QR, United Kingdom}
\affil[2]{Department of Energy Conversion and Storage, Technical University of Denmark, \protect\\ Kgs.~Lyngby 2800, Denmark}

\date{}
\maketitle
\thispagestyle{empty}

\begin{abstract}
The chalcogenide perovskite material \ce{BaZrS3} is of growing interest for emerging thin-film photovoltaics. Here we show how machine-learning-driven modelling can be used to describe the material's amorphous precursor as well as polycrystalline structures with complex grain boundaries. Using a bespoke machine-learned interatomic potential (MLIP) model for \ce{BaZrS3}, we study the atomic-scale structure of the amorphous phase, quantify grain-boundary formation energies, and create realistic-scale polycrystalline structural models which can be compared to experimental data. Beyond \ce{BaZrS3}, our work exemplifies the increasingly central role of MLIPs in materials chemistry and marks a step towards realistic device-scale simulations of materials that are gaining momentum in the fields of photovoltaics and photocatalysis.
\end{abstract}

\clearpage

\setstretch{1.5}

\section*{Introduction}

In the search for new, sustainable photoabsorbers, sulfide-based chalcogenide perovskite materials have emerged as attractive lead-free candidates.\cite{gupta_environmentally_2020} 
However, while oxide and halide perovskites have defined much of the progress in photovoltaics and related fields, chalcogenide perovskites have only more recently begun to be explored. Among the latter, \ce{BaZrS3} presents optical absorption matching or even surpassing those of halide perovskites and GaAs,\cite{nishigaki_extraordinary_2020} competitive charge carries lifetimes, and improved stability to environmental factors compared to other perovskite materials.\cite{comparotto_chalcogenide_2020, yu_chalcogenide_2021, thakur_recent_2023}  Thin films of \ce{BaZrS3} can be synthesised from earth-abundant and non-toxic elements: by sulfidation of Ba--Zr--O precursors \cite{wei_realization_2020, marquez_bazrs3_2021, yu_chalcogenide_2021} or by directly depositing sulfide species using pulsed laser deposition,\cite{surendran_epitaxial_2021} molecular beam epitaxy,\cite{sadeghi_making_2021} or sputtering.\cite{comparotto_chalcogenide_2020} Most methods involve the deposition of amorphous precursors that require temperatures of $\approx$ 900 $\degree$C to crystallise.
Their growth and subsequent crystallisation has  been followed experimentally using X-ray diffraction (XRD) or X-ray spectroscopy techniques. \cite{mukherjee_interplay_2023,ramanandan_understanding_2023,riva_electronic_2024}

Given the rapidly growing interest in \ce{BaZrS3}, computational methods are increasingly used to complement experimental studies of this material. Density-functional theory (DFT) and phonon computations were employed to map out the thermodynamic conditions under which \ce{BaZrS3} films might form and which surface termination is expected to be the most stable.\cite{kayastha_first-principles_2024, osei-agyemang_understanding_2023} To reach beyond the system-size limits of DFT-based methods, 
machine-learned interatomic potentials (MLIPs) have now been applied to many functional materials, \cite{behler_first_2017, deringer_machine_2019, friederich_machine-learned_2021} including halide perovskites.\cite{jinnouchi_phase_2019, braeckevelt_accurately_2022, fransson_phase_2023, baldwin_dynamic_2023} 
The chalcogenide alternatives, {\em viz.} \ce{BaZrS3} and homologous compounds, were recently studied in a comprehensive work using ML-accelerated molecular dynamics (MD).\cite{jaykhedkar_how_2023} These studies have typically focused on the crystalline material \cite{fransson_phase_2023, jaykhedkar_how_2023} and the formation of other phases, such as the binary crystals or 2D Ruddlesden--Popper structures.\cite{kayastha_first-principles_2024} To validate ML-accelerated MD, Kayastha {\em et~al.}\ compared simulated XRD patterns for MD-generated \ce{BaZrS3} structures with experimental XRD patterns.\cite{kayastha_octahedral_2025} However, these simulations also were focused on ordered unit cells, corresponding to single-crystalline samples.

\begin{figure*}[t]
\centering
  \includegraphics[width=0.85\textwidth]{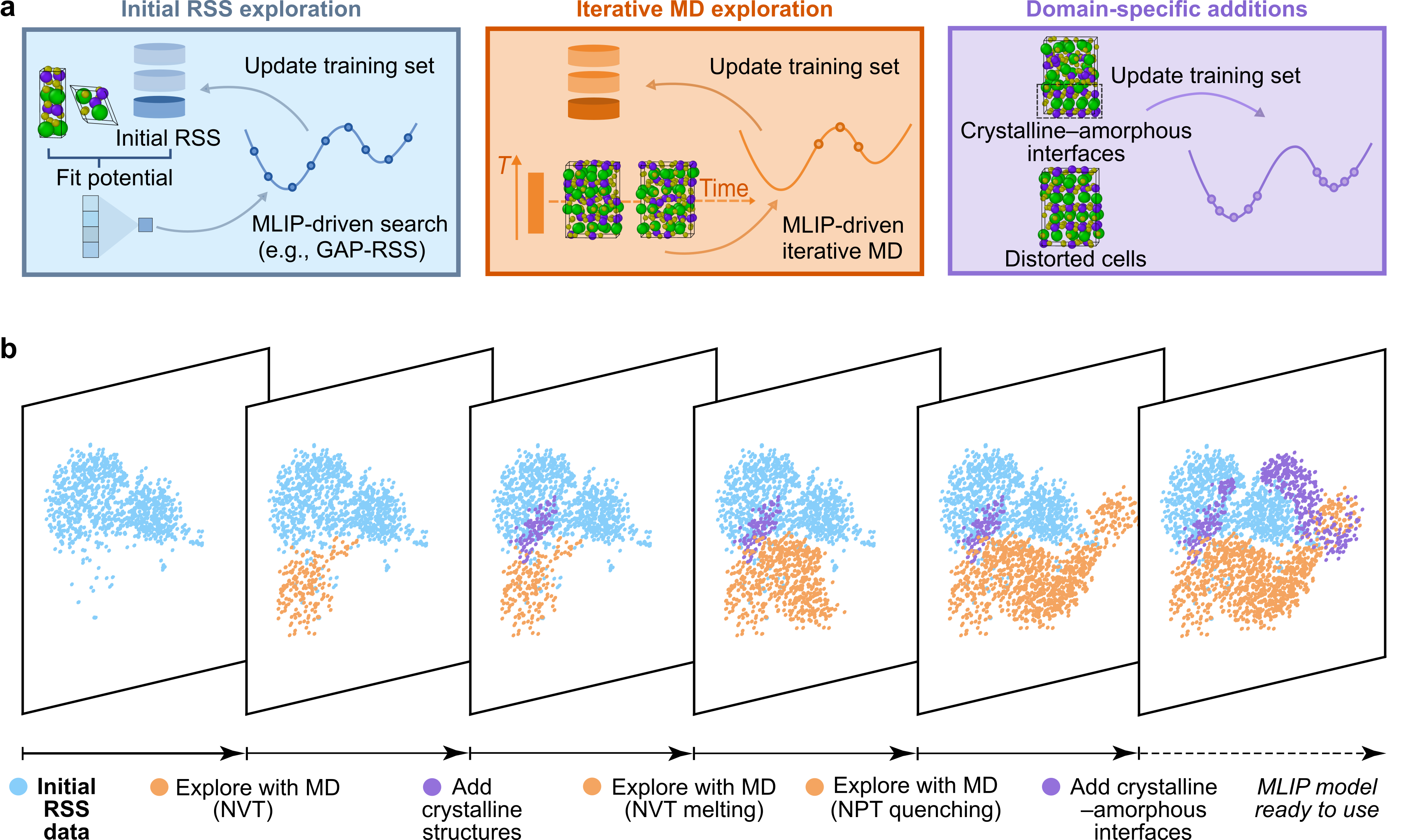}
  \caption{A machine-learned interatomic potential for crystalline and amorphous \ce{BaZrS3}.
  (\textbf{a}) Schematics of the different approaches used in training dataset construction, showing examples of the different configuration types sampled. Structural images were created using OVITO.\cite{stukowski_visualization_2009}
  (\textbf{b}) Evolution of the training dataset, visualised in a style similar to Ref.~\citenum{el-machachi_accelerated_2024}. Each slice provides a two-dimensional representation (using the UMAP algorithm\cite{mcinnes_umap_2018}) of the relevant configurational space, showing the reference training dataset of the ML potential, distributed based on the structures' average atomic-environment similarity. The latter is quantified using the SOAP kernel similarity metric \cite{bartok_representing_2013} with a cut-off radius of 5 \AA{} and smoothness of 0.75 \AA{} (see Supplementary Information for more details): the distances between points in this two-dimensional space therefore reflect the structural (dis-) similarity between entries of the training dataset.}
  \label{fgr:example}
\end{figure*}

This limitation is more generally a current research challenge in modelling perovskite solar-cell materials: experimentally synthesised materials are usually polycrystalline, and fully realistic simulations would therefore need to involve structural models representing individual grains, with sizes typically on the order of tens or hundreds of nanometres. \cite{mukherjee_interplay_2023, riva_electronic_2024} 
A single-crystalline structural model will therefore likely not suffice to fully understand the structure and properties of \ce{BaZrS3} films. 
We have recently reported very-large-scale atomistic models of functional materials, including phase-change materials for data storage \cite{zhou_device-scale_2023} and amorphous silicon which is relevant to solar cells. \cite{Morrow2024, Rosset2025} It would seem highly beneficial to achieve this type of realism for chemically complex, perovskite-type photoabsorber materials as well.

Here, we introduce a machine-learned interatomic potential (MLIP) model for ordered and disordered forms of \ce{BaZrS3}, based on the atomic cluster expansion (ACE) framework.\cite{drautz_atomic_2019, lysogorskiy_performant_2021}  
For training, we employ a combination of {\em de novo} \cite{deringer_data-driven_2018, bernstein_novo_2019} and domain-specific iterative training (Fig.~\ref{fgr:example}), aiming for the final dataset to capture the structural complexities of \ce{BaZrS3}.
We show how ML-driven simulations can describe three scenarios relevant to experimental studies: (i) the amorphous precursor; (ii) large-scale grain boundaries; and (iii) polycrystalline \ce{BaZrS3} structures. This way, ML-driven simulations can corroborate experimental observations regarding the atomistic structure of this material and provide insights that experiments on their own can not. 
Beyond their application to \ce{BaZrS3}, we expect that ML-driven approaches for simulating polycrystalline structures---from precursors to individual grains---can more broadly accelerate computational studies of diverse polycrystalline solar-cell materials. 

\section*{Methodology}

The choice of data used to train MLIP models is now a central consideration in the field,\cite{ben_mahmoud_data_2024} and different approaches to dataset building have been discussed in the literature.\cite{deringer_gaussian_2021, kulichenko_data_2024}
Here, we begin with random structure searching (RSS)\cite{PhysRevLett.97.045504, pickard_ab_2011} using an iterative protocol similar to Refs.~\citenum{deringer_data-driven_2018} and  \citenum{bernstein_novo_2019}, whereby an initial MLIP is trained on randomised structures and then used to sequentially drive the RSS exploration (Fig.~\ref{fgr:example}a). It was previously shown that RSS can sample the complex atomistic environments relevant to grain boundaries and interfaces by generating diverse starting structures.\cite{schusteritsch_predicting_2014} High-temperature structures obtained from ML-driven MD, as well as crystalline--amorphous interfaces were subsequently added (Supplementary Information). As shown in a series of ``structure maps'' in Fig.~\ref{fgr:example}b, the structures generated at different temperatures using iterative MD (shown as orange points in Fig.~\ref{fgr:example}b) sample different regions of configurational space compared to the initial RSS dataset (light blue). Furthermore, our ``domain-specific'' additions to the dataset (purple), such as small-scale structural models representing crystalline--amorphous interfaces, include structures in-between disordered, high-temperature snapshots from MD and crystalline structures. Domain-specific training data such as interface structures have been used before to help describe crystallisation processes in Ge--Sb--Te memory materials, for example. \cite{zhou_device-scale_2023}

We used different MLIP fitting approaches as part of the development of the final model. Initially, the Gaussian Approximation Potential (GAP)\cite{bartok_gaussian_2010} framework was used because of its data efficiency: it allowed us to generate a stable initial potential with relatively few training data points (90 initial RSS structures, with a further 899 structures obtained from {\em de novo} GAP-RSS exploration\cite{deringer_data-driven_2018, bernstein_novo_2019}). Once a larger dataset had been built by iterative training, the computationally efficient ACE framework as implemented in \texttt{pacemaker}  was used to fit a faster MLIP model to that dataset.\cite{lysogorskiy_performant_2021, zhou_full-cycle_2025}

The final ACE model was obtained by iterative training until it could reliably generate a structural model for the amorphous phase using an MD melt--quench protocol in the NPT ensemble (see Supplementary Information for further details).  
Two model versions were fitted to the final dataset: the first by filtering out structures with high DFT energies (> 1 eV/atom), indicative of very close contacts between atoms or high-energy RSS structures, and the second using the full reference dataset (Supplementary Information). The first version was used to generate all the quantitative data presented, as it achieved good accuracy on the structures relevant in the present study, showing an energy root-mean-square error (RMSE) of 13.9 meV/atom relative to DFT results using the PBEsol functional.\cite{perdew_restoring_2008} The second version, which includes higher-energy dimers and random structures, is less accurate (energy RMSE: 23.1 meV/atom), but it did not fail when handling structures with closer contacts between atoms, and therefore it was used to relax the polycrystalline structures with randomly-oriented grains. (We consider a simulation to have ``failed'' if, during the MD run, atoms come closer to each other than 1 \r{A}, collapsing the structure, or if atoms are lost during the simulation.)
Details of numerical errors are provided in Figs.\ S1 and S2 in the Supplementary Information.

\section*{Results and discussion}

We describe the computational modelling of \ce{BaZrS3} in the same sequence as would be relevant to experimental synthesis and characterisation. First, we use the MLIP model to simulate the amorphous phase, corresponding to precursor phases that are deposited in experiments. \cite{comparotto_chalcogenide_2020, yu_chalcogenide_2021} Second, we validate the model for grain boundaries, which need to be accurately described so that the model can be applied to polycrystalline samples. With both of these aspects available, we finally apply the model to simulating structures with different grain sizes, providing a direct connection to experimental scattering data.

\begin{figure*}[t]
\centering
  \includegraphics[width=\textwidth]{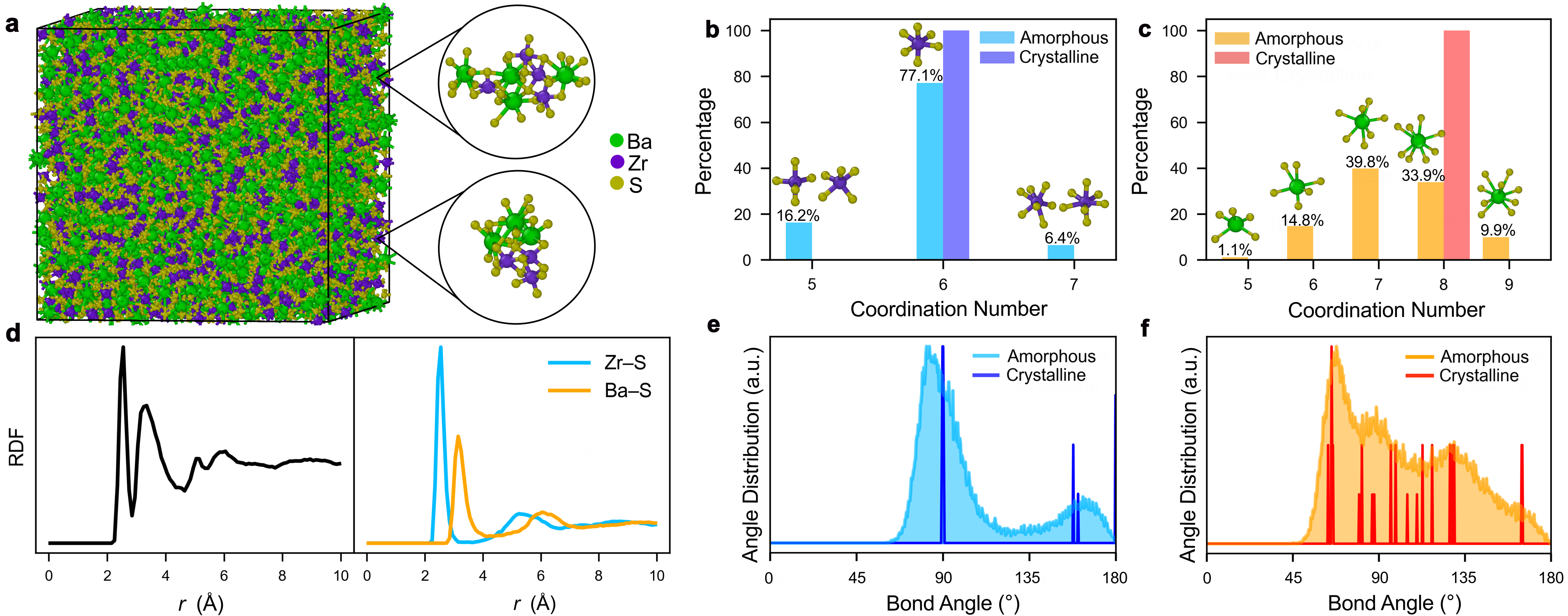}
  \caption{Amorphous \ce{BaZrS3}.
  (\textbf{a}) The atomistic structure of the simulated amorphous phase (10,240-atoms) and manually-chosen close-ups showing the range of geometries and connectivity types between the coordination polyhedra of the A- and B-site cations. 
   (\textbf{b}) Histogram plot showing the distribution of coordination numbers around Zr atoms in the amorphous and crystalline phases, respectively. (Note that environments with CN $< 0.5\%$ are not visualised in the histogram, but Table S3 presents full details of the coordination numbers.) (\textbf{c}) Same but for Ba atoms. (\textbf{d}) Total and partial RDFs for the Zr--S and Ba--S interatomic distances in the 10,240-atom simulated amorphous phase. (\textbf{e}) ADF plot showing the S--Zr--S bond-angle distribution in the amorphous phase, compared to the crystalline structure. (\textbf{f}) Same but for S--Ba--S bond angles. The radial cut-off used for calculating coordination numbers and ADF plots was set to 3.1 \r{A} in the case of the Zr environments and to 3.8 \r{A} in the case of Ba. }
  \label{fgr:amorph}
\end{figure*} 

\subsection*{Amorphous \ce{BaZrS3}}

The amorphous structure simulated using the MLIP model is shown in Fig.~\ref{fgr:amorph}a. In this structure, many of the Zr atoms still have a (defective) octahedral coordination by S, similar to the crystalline structure; however, the ZrS$_x$ polyhedra lack long-range order. The geometry of the different Zr coordination environments, which also present undercoordinated and a few overcoordinated Zr atoms, and those of the BaS$_x$ polyhedra, are shown in the histogram plots in Fig.~\ref{fgr:amorph}b--c, respectively. We show examples of the coordination polyhedra for each coordination number (CN). A wider distribution of CNs is observed for Ba, and in this case, also, we observe a pronounced undercoordination of the cation, which has an expected CN of 8 in crystalline \ce{BaZrS3} (due to the orthorhombic distortion of its perovskite-like structure, and compared to CN = 12 for the cubic archetype).

The local ordering of S atoms around the A- and B-site cations can be observed in the radial distribution function (RDF) of the quenched amorphous structure (Fig.~\ref{fgr:amorph}d, left), which shows well-defined peaks in the short-range region, at distances roughly below 4~\r{A}. These two main peaks correspond to the Zr--S and Ba--S interatomic distances, as confirmed by the partial RDF plots (Fig.~\ref{fgr:amorph}d, right). 
The RDF data obtained from atomistic simulations can be compared with experimental extended X-ray absorption fine structure spectroscopy (EXAFS) results,\cite{mukherjee_interplay_2023} which probe the local structure around Zr atoms in the amorphous phase. Although the techniques are different, both the EXAFS data and our simulations qualitatively indicate under-coordinated Zr environments in amorphous \ce{BaZrS3} compared to its crystalline counterpart. Specifically, analysis of the EXAFS data yielded a Zr–S bond length of 2.593 \AA{} and coordination number (CN) of 5.2 (see Ref.~\citenum{mukherjee_interplay_2023} for details), while our simulations yield an average bond length of 2.575 \AA{} and CN of 5.9 (determined using a 3.1 \AA{} cut-off in OVITO\cite{stukowski_visualization_2009}). 
Furthermore, the presence of B-site cation halide fragments, some maintaining a similar octahedral geometry to the crystalline phase, is in agreement with experimentally reported structural details of the amorphous phases derived from other materials adopting the perovskite structure.\cite{wachtel_quasi-amorphous_2010, rigter_passivation_2021, mukherjee_interplay_2023} Preserved local bonding units of \ce{TiO6} connected in a random network via apex-, edge-, and face-sharing octahedra have also been observed in the amorphous phases of \ce{BaTiO3}\cite{frenkel_microscopic_2005, ehre_structural_2007} and \ce{SrTiO3}.\cite{frenkel_origin_2007} 

The expected bulk density of amorphous \ce{BaZrO3} phases was reported to be in the range of 80--85\% of the crystalline density (Ref.~\citenum{ramanandan_understanding_2023}). In our ML-driven NPT simulation of amorphous \ce{BaZrS3}, the observed density was 3.94 g/cm$^3$, which is roughly 92.5\% of the crystalline density, in qualitative agreement with the lower density observed in the related oxide compound. 
The average bond length of the 6-fold-coordinated Zr atoms in the amorphous phase, with a value of 2.58 \AA{}, is similar to the expected bond length of 2.55--2.56 \AA{} in the ZrS$_6$ octahedra of the crystalline phase. In the case of the 5- and 7-coordinated Zr--S environments, the bonds are slightly compressed or elongated compared to the crystalline phase (2.50 \AA{} and 2.66 \AA{}, respectively). The Ba--S bonds vary in length within a similar range to that observed in the crystalline phase, around 3.0--3.4 \AA{}; however, as also observed in the partial RDF peak, the coordination can vary more than in the case of the Zr environment. 

The relationship between the geometry of the cation environments in the amorphous and crystalline forms of \ce{BaZrS3} can be observed in the angle distribution function (ADF) plots for the S--Zr--S and S--Ba--S bond angles (Fig.~\ref{fgr:amorph}e and Fig.~\ref{fgr:amorph}f, respectively). The main ADF peaks in the case of Zr are clearly distributed around the values expected for the octahedral crystalline environment, specifically 90$\degree$ angles between equatorial and axial Zr--S bonds, 180$\degree$ between the axial Zr--S bonds, and about 150$\degree$ for the bonds connecting the octahedra in the orthorhombic crystal structure. The distribution is harder to assess for the S--Ba--S bond angles, both due to the wider range of bond angles observed for higher coordination geometries, and due to the greater variety of CNs present in the amorphous phase. Given the undercoordination of the cations compared to the crystalline phase, there is an observed increase in close S--S contacts in the amorphous structure (Fig.~S3).

We note that structural properties of the amorphous phase are of interest not only to provide insight into the atomic environments found in the as-deposited precursor to the polycrystalline material, but additionally to reveal possible structure--property relationships in amorphous or surface-amorphised perovskites which have shown good performance as electro- or photoelectrocatalysts in the case of oxide perovskite materials.\cite{sun_perovskite_2024}
The presence of dangling bonds from undercoordinated atoms has been suggested as a possible reason for the efficiency of perovskites with amorphised surfaces during electrochemical processes.\cite{sun_perovskite_2024} Further experimental work could determine whether electrocatalytic surface reconstruction occurs in \ce{BaZrS3} and therefore whether this could explain the performance of the material in electrocatalysed oxygen or hydrogen evolution reactions, where the reactant binds to undercoordinated Zr sites.\cite{yasmin_solar_2023, humphrey_hydrogen_2024}
In the long run, knowledge of the coordination environment in the amorphous precursor phase might aid in the development of higher-quality crystallised materials with fewer defects.

\begin{figure*}[t]
\centering
  \includegraphics[width=\textwidth]{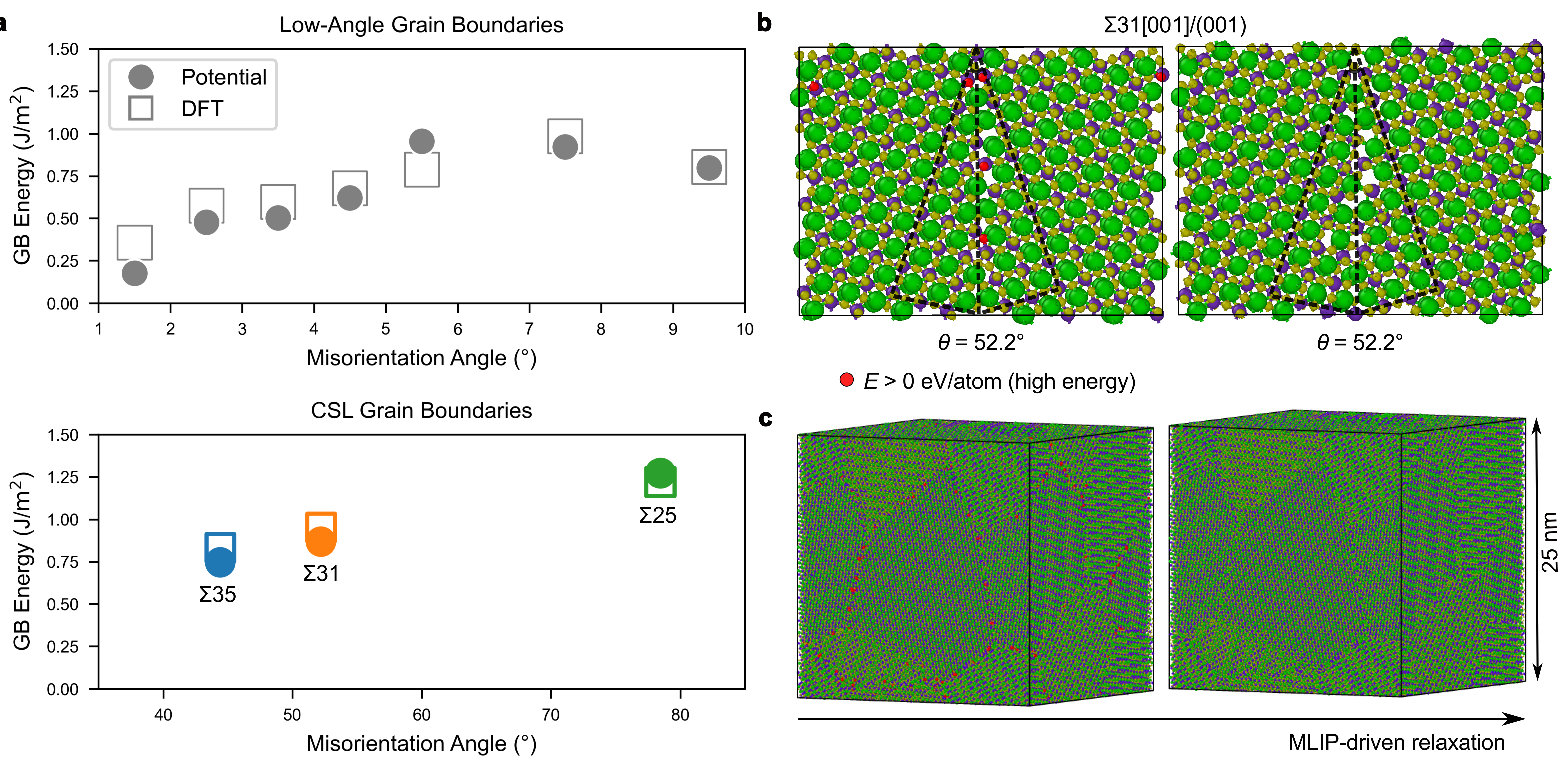}
  \caption{Grain boundaries.
  (\textbf{a}) The grain-boundary (GB) energy prediction for relaxed GB structures of \ce{BaZrS3} using the MLIP model compared with the single-point energies predicted by DFT at the PBEsol level of theory. (\textbf{b}) Optimisation of the $\Sigma$31[001]/(001) GB geometry, where high-energy atoms with energies higher than 1 eV/atom (red) are relaxed to a lower-energy structure using the ML potential. (\textbf{c}) Relaxation of a 615,214-atom 3D polycrystalline structure with six randomly-oriented grains.}
  \label{fgr:GB}
\end{figure*}

\subsection*{Grain boundaries}

Crucially, the ML potential also allows the modelling of interfaces, such as those observed at grain boundaries (GBs). To date, the atomistic structure of the boundary region and the associated GB formation energies in \ce{BaZrS3} have remained unexplored. Therefore, we use the misorientation angles predicted by coincidence site lattice (CSL) theory applied to an orthorhombic system \cite{king_geometrical_1993} to construct models of the anticipated stable structures for higher misorientation angles around the (001) axis, in the  lattice plane defined by the \textit{a} and \textit{b} axes. Additionally, we test the potential's performance on low-angle GBs, choosing misorientation angles up to 10$\degree$. The MLIP was used to relax the positions of atoms in the boundary region to obtain structures with GB energies in good agreement with DFT results (Fig.~\ref{fgr:GB}a). The GB formation energy, $E_{\textnormal{GBf}}$, was calculated as
\[E_{\textnormal{GBf}}=\frac{E_{\textnormal{GB}}-E_{\textnormal{bulk}}}{2A},
\]
where $E_{\textnormal{GB}}$ is the energy of the modelled system containing two identical grain boundaries, $E_{\textnormal{bulk}}$ is the energy of the same supercell without grain boundaries (the bulk single-crystal system), and $A$ is the area of the grain-boundary region. 

For most GB systems studied, the MLIP's prediction has a minimum accuracy of 0.1 J~m$^{-2}$ relative to DFT values. This is within 0.07 J m$^{-2}$ of the errors obtained in a study by Ito {\em et~al.}\ for an ML potential specialised on grain-boundary structures.\cite{ito_machine_2024} The maximum error is obtained for the GB with a misorientation angle of 1.5$\degree$, which is 0.18 J m$^{-2}$ off the DFT value. Overall, the potential is also able to capture the relative stabilities of the different GB systems, with the exception of the GB with an angle of 5.5$\degree$, where the energy prediction incorrectly identifies it as higher in energy than its relative ground-truth value. These errors could be addressed by adding the structures of interest to the dataset or changing the weightings of relevant configuration types such that the potential is more specialised on the region of interest (Fig.~S4). As noted, it has been previously shown that ML potentials can be specifically trained on different GB structures.\cite{ito_machine_2024}$^,$\cite{yokoi_accurate_2022} However, such targeted training is beyond the scope of the current work. 

An example of the successful relaxation of the expected CSL $\Sigma$31[001]/(001) GB system is shown in Fig.~\ref{fgr:GB}b. The relaxation of an approximately 1,000-atom GB structure, at the limit of what is achievable using DFT methods, is achieved within seconds using the MLIP, while systems of up to hundreds of thousands of atoms can be successfully studied and relaxed within a few minutes on a 128-core CPU compute node.

\subsection*{Polycrystalline structures}

As discussed above, MLIPs allow the relaxation of much larger and more realistic systems which are inaccessible to \textit{ab initio} methods. Going beyond a simple grain-boundary system, we created 3D polycrystalline models of up to 600,000 atoms (Fig.~\ref{fgr:GB}c).  The polycrystalline unit cell with 6 randomly-oriented grains was generated by Voronoi tessellation in Atomsk\cite{hirel_atomsk_2015} and relaxed with the potential trained on the full dataset, to avoid unphysical close contacts between atoms (see Supplementary Information for details). The extrapolation grade based on the D-optimality algorithm is an established method for measuring the uncertainty of an ACE ML potential in a particular region of the configurational space being modelled.\cite{lysogorskiy_active_2023, ito_machine_2024} We found that the MLIP is able to relax the polycrystalline unit cell, resulting in a structure with extrapolation grades up to a maximum value of 2, which is considered accurate, as described in Ref.~\citenum{lysogorskiy_active_2023} (see Fig.~S5). This result suggests that the potential can predict the atomistic structure with relatively high certainty, including in the boundary region. Given the potential's performance for modelling the amorphous phase, its transferability to the polycrystalline system is unsurprising, as we found that in the polycrystalline relaxed structure, the coordination around the Zr atom is similar to that found in the amorphous phase, but with a shorter Zr--S bond length (CN$_\text{Zr}$ = 5.87 and $d_{\text{Zr}-\text{S}}$ = 2.54 \r{A}; see also Fig.~S6). Thus, the undercoordinated Zr environment suggests that the relaxed grain boundaries are S-deficient (Fig.~S7).

\begin{figure*}[t]
\centering
  \includegraphics[width=\textwidth]{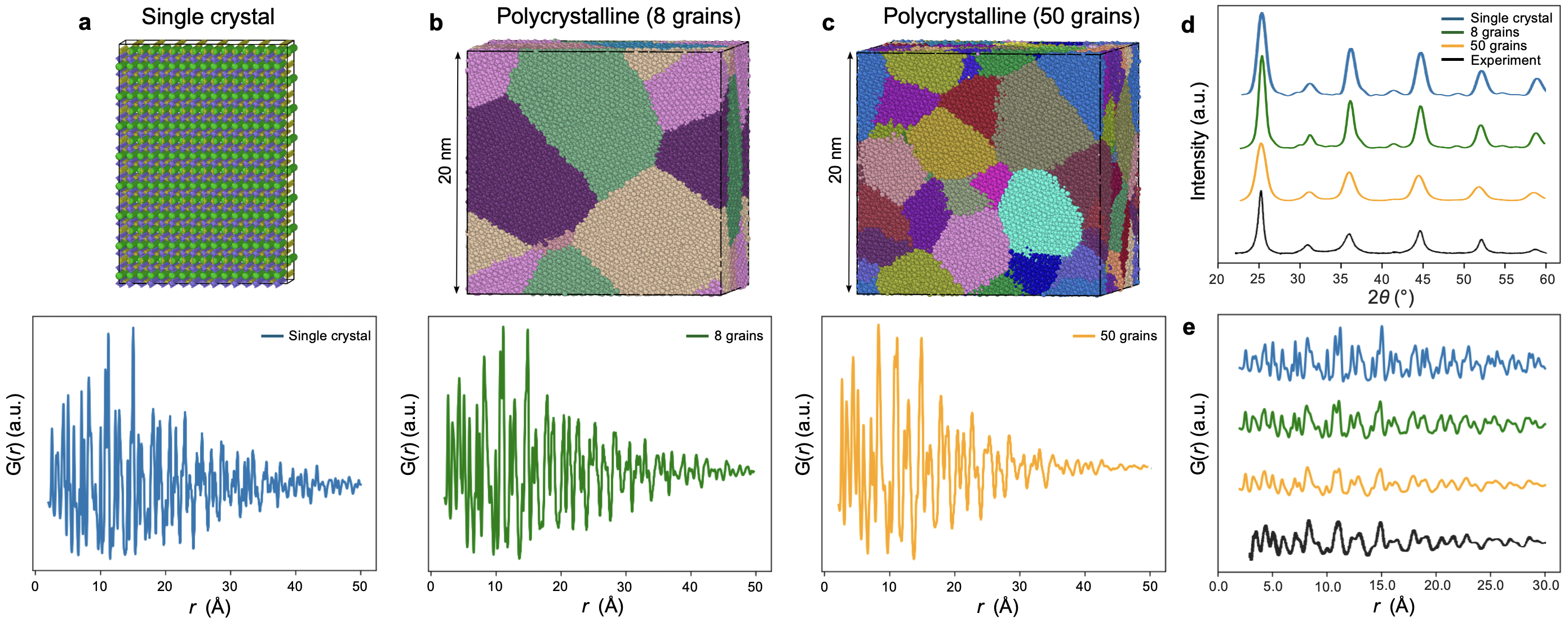}
  \caption{Effect of grain size on the X-ray diffraction (XRD) patterns and pair distribution functions (PDFs) of crystalline \ce{BaZrS3}. (\textbf{a}) A single-crystal 10,240-atom model (top) and its simulated PDF [$G(r)$; bottom]. (\textbf{b}) A polycrystalline model with a simulation box side length of 20 nm containing 8 grains (313,449 atoms). (\textbf{c}) A polycrystalline model with a simulation box side length of 20 nm containing 50 grains (312,370 atoms). (\textbf{d}--\textbf{e}) Comparison between the XRD patterns and PDFs, respectively, for the simulated models and the experimental data from Ref.~\citenum{zilevu_solution-phase_2022}.
  For simulating the scattering data, we employed the DebyeCalculator default parameters, except for the intensity data shown in panel (d) where $Q_\textnormal{min}$ and $Q_\textnormal{max}$ were set to 1.61 \AA{}$^{-1}$ and 5.25 \AA{}$^{-1}$, respectively, assuming the use of Cu K-$\alpha$1 radiation for the conversion to 2$\theta$. }
  \label{fgr:GB-size}
\end{figure*} 

While the potential can efficiently relax systems exceeding 600,000 atoms, calculating scattering patterns of discrete structures remains computationally demanding,\cite{Johansen2024} and so we focused our study on polycrystalline unit cells with a side length of 20 nm ($\approx$ 310,000 atoms). Figure~\ref{fgr:GB-size} illustrates the effect that polycrystallinity has on the simulated XRD pattern and pair distribution function (PDF) of \ce{BaZrS3}, which were generated using the DebyeCalculator package (Ref.~\citenum{Johansen2024}). Going from a single-crystal model shown in Fig.~\ref{fgr:GB-size}a to a polycrystalline model with 8 grains and 50 grains in Figs.~\ref{fgr:GB-size}b and \ref{fgr:GB-size}c, respectively, the peaks in the simulated XRD patterns broaden---an effect primarily attributable to size variation: the size of the grains decreases when their number increases in a simulation box of constant size, as expected from applying the Scherrer equation in XRD experiments. This broadening causes the smaller, secondary peaks observed in the crystalline pattern to gradually become unresolved for the systems with a larger number of grains, as shown in Fig.~S8, which compares the simulated partial XRD patterns of single-crystal and polycrystalline \ce{BaZrS3} models. 

In the PDF, size effects manifest as a more rapid dampening of the signal at high $r$ values in the model with a large number of grains relative to the polycrystalline model with fewer, larger grains and to that of the single-crystal model. Additionally, the local structure of each model varies slightly, presumably due to differences in the grain-boundary fraction. In Fig.~\ref{fgr:GB-size}d--e, the simulated data are compared to experimental data from Ref.~\citenum{zilevu_solution-phase_2022}. The experimental XRD and PDF patterns are described best by the polycrystalline 50-grains model, consistent with the nanoparticulate nature of the material described in Ref.~\citenum{zilevu_solution-phase_2022}, which presents a solution-phase synthesis of plate-like, aggregated \ce{BaZrS3} nanoparticles. Similar results were also reported in Ref.~\citenum{yang_low-temperature_2022}: a low-temperature synthesis of \ce{BaZrS3} nanoparticles with grain sizes of 3--5 nm. A related trend in the experimental XRD patterns can be observed when samples are progressively annealed at higher temperatures, whereby the crystal size increases.\cite{comparotto_chalcogenide_2020}

\section*{Conclusions and outlook}

Our study has introduced an ML-based interatomic potential model for disordered and polycrystalline \ce{BaZrS3} and shown how it can be used to simulate multiple types of structures that are relevant for this emerging functional material. We addressed questions related to the structural and physical properties of \ce{BaZrS3} across different stages of the experiment, from the deposition of the amorphous precursor phase to the study of polycrystalline systems with different grain sizes. These simulations can be directly compared with experimental results, such as XRD patterns, PDFs, or local bonding information obtained from EXAFS. Furthermore, the potential is capable of relaxing grain-boundary structures and estimating their formation energies at a low computational cost for large-scale structural models: simulations on the order of hundreds of thousands of atoms can be achieved in less than a day on a single 64-core compute node. 

The present study is an early step towards the realistic modelling of thin-film photovoltaic materials. 
In the future, as an extension to this work, the modelling of surface structures, or amorphised surfaces, could provide further insights into the material's functionality. 
More generally, our work shows how polycrystalline perovskite materials can now be modelled with atomistic machine learning, having in view the realism that has already been achieved for other technologically relevant systems. \cite{zhou_device-scale_2023, Chen2025} The present study thus lays the groundwork for such studies involving more complex photovoltaic materials, including mixed-cation and anion perovskites, with hybrid organic--inorganic components, and for starting to approach their modelling at the length scale of real devices.

\clearpage

\setstretch{1.1}

\section*{Acknowledgements}

We thank D.\ F.\ Thomas du Toit and Y.\ Zhou for helpful discussions.
L.-B.P. acknowledges funding from the EPSRC Centre for Doctoral Training in Inorganic Chemistry for Future Manufacturing (OxICFM), EP/S023828/1. 
This work was supported by UK Research and Innovation [grant numbers EP/X016188/1 and EP/Z000343/1].
The work presented in this article is supported by Novo Nordisk Foundation grant NNF23OC0081359.
We are grateful for computational support from the UK national high performance computing service, ARCHER2, for which access was obtained via the UKCP consortium and funded by EPSRC grant ref EP/X035891/1.  

\section*{Data Availability Statement}

Data supporting this work are available at \href{https://github.com/BiancaPasca/polycrystalline-BaZrS3}{https://github.com/BiancaPasca/polycrystalline-BaZrS3}.

\providecommand*{\mcitethebibliography}{\thebibliography}
\csname @ifundefined\endcsname{endmcitethebibliography}
{\let\endmcitethebibliography\endthebibliography}{}

\end{document}


\title{\Large {\bf Supplementary Information for}\\[4mm] ``Machine-learning-driven modelling of amorphous and  polycrystalline \ce{BaZrS3}''}

\author[1]{Laura-Bianca Pa\c{s}ca}
\author[1]{Yuanbin Liu}
\author[1,2]{Andy S. Anker}
\author[1]{Ludmilla Steier}
\author[1]{Volker L. Deringer\thanks{volker.deringer@chem.ox.ac.uk}}

\affil[1]{Inorganic Chemistry Laboratory, Department of Chemistry, University of Oxford, \protect\\ Oxford OX1 3QR, United Kingdom}
\affil[2]{Department of Energy Conversion and Storage, Technical University of Denmark, \protect\\ Kgs.~Lyngby 2800, Denmark}

\date{}

\maketitle

\thispagestyle{empty}

\clearpage
\setstretch{1.1}

\section*{Computational methods}

\subsection*{Training data}

Table S1 shows the composition of the training dataset. The structures were split in a 9:1 ratio between the training and test datasets in each different iteration.

\begin{itemize}
    \item \textbf{Free atoms and dimers.} The isolated atoms (Ba, Zr, S), as well as the six different pairs of dimers (Ba--Ba, Zr--Zr, S--S, Ba--Zr, Ba--S, Zr--S) in vacuum were included using 20 × 20 × 20 \r{A}$^3$ simulation cells. The dimers were scaled to sample distances between 1 and 5 \r{A} and only those with reference forces that did not exceed 100 eV/\r{A} were added to the dataset after labelling.
    \item \textbf{``iter0'': Random Structure Search (RSS) data.} The structures in this iteration were generated using the {\tt buildcell} code from the AIRSS package.\cite{pickard_ab_2011} Different constraints were used to produce a diverse set of 10,000 random structures containing between 1–10 formula units, with a target volume in the range of 20–25 \r{A}$^3$/atom and a constraint on the minimum separation between atoms ({\tt \#MINSEP=2.6 Ba-S=2.0 Zr-S=1.8 Ba-Zr=2.4 S-S=2.2}). These structures were then labelled using a Smooth Overlap of Atomic Positions (SOAP) similarity descriptor, \cite{bartok_representing_2013} as detailed in the \hyperref[subsec:hyp]{Model hyperparameters} section. Based on their SOAP vectors, the 100 most diverse structures were picked using the leverage-score CUR algorithm, \cite{mahoney_cur_2009} and finally these were split in a 9:1 ratio between the training (``iter0'') and test datasets.
    \item \textbf{``iter1'': GAP-driven RSS structures.} More random structures were generated using a GAP-driven RSS approach \cite{bartok_gaussian_2010, bernstein_novo_2019}, whereby the previous iteration of the GAP potential is used to minimise the enthalpy of newly-generated structures obtained using the {\tt buildcell} code. A combination of Boltzmann-biased flat histogram and CUR decomposition are then used to select the 100 most diverse structures with lower energies, as described in a previous work.\cite{bernstein_novo_2019} In terms of the biasing factor, which determines the probability of structure selection, the temperature is decreased from 0.5 eV to 0.1 eV across 10 different iterations, with a decrease of 0.1 eV every 2 iterations. One structure with large forces ($>$ 80 eV/\r{A}) was discarded.
    \item \textbf{``iter2'': NVT MD starting from RSS structures.} The MD simulations were run using LAMMPS\cite{thompson_lammps_2022} with a timestep of 1 fs and using the Nos\'e{}--Hoover thermostat.
    Three 100-atom ({\tt \#NFORM=20}) cubic cells of different densities (2.95 g/cm$^3$ , 3.58 g/cm$^3$ and 4.22 g/cm$^3$) were generated using RSS as described above, additionally setting a constraint on the unit-cell parameter to control the volume of the cells. GAP-driven MD simulations in the NVT ensemble were ran for each different cell at 1,500 K and 2,500 K  to create the ``iter2-1'' and ``iter2-2'' datasets, respectively. The structures were first equilibrated at 300 K for 10 ps, then the temperature was brought to the corresponding maximum value at a rate of 10$^{14}$ K/s, followed by an equilibration at that temperature for 10 ps. Finally, the structure was held at high temperature for another 10 ps, and the different snapshots were selected from this point on in the trajectory.
    From the collected snapshots, 30, 40 and 30 structures were selected by CUR for the 2.95 g/cm$^3$, 3.58 g/cm$^3$ and 4.22 g/cm$^3$ cells, respectively, then added to the training (``iter2-1'') and test datasets in a 9:1 ratio. 
    \item \textbf{Crystalline structures.} The stable ternary crystalline structures with the Ba--Zr--S composition reported in the Materials Project database\cite{jain_commentary_2013} were all added to the training set. The crystalline data were augmented by rattling, and distorting the angles and volumes of the unit cells.
    \item \textbf{``iter3'': NVT melted structures.} Starting from a 180-atom supercell of the relaxed orthorhombic \ce{BaZrS3} structure ($Pnma$) with the original crystalline volume, as well as three cells with different scaled volumes (1.05$\times$, 1.10$\times$, and 1.20$\times$ the crystalline volume), NVT trajectories were generated at 3,500 K, 4,000 K, and 4,500 K. Snapshots with unphysical atomic distances (< 1 \r{A}) were discarded. From each of the 12 trajectories (3 temperatures $\times$ 4 unit cell densities), 25 snapshots were selected, and the dataset was split into training (``iter3'') and test sets in a 9:1 ratio.
    \item \textbf{``iter4'': NPT trajectories using ACE.} The \textit{pacemaker} implementation of ACE \cite{lysogorskiy_performant_2021} was used to train a potential based on all the training data up to and including ``iter3''. The first two iterations, ``iter4-1'' and ``iter4-2'', included snapshots from melt-quench NVT trajectories starting with a 180-atom crystalline \ce{BaZrS3} supercell. For ``iter4-1'', the structure was equilibrated at 300 K for 20 ps, heated to 3,000 K at a rate of 10$^{14}$ K/s, and then cooled to 1,500 K (near the melting point) at the same rate. Quenching from 1,500 K to 300 K was performed at four different rates (10$^{14}$, 2.5$\times$10$^{14}$, 5$\times$10$^{14}$, and 10$^{15}$ K/s), with 25 snapshots evenly selected from each trajectory.
    For ``iter4-2'', the process was repeated using the updated ACE potential, with quench rates of 10$^{13}$ and 10$^{12}$ K/s, and 100 evenly-spaced snapshots selected from the two trajectories.
    In ``iter4-3'', NPT simulations with 360-atom cells were run at quench rates of 10$^{12}$, 10$^{13}$, and 10$^{14}$ K/s. From these trajectories, 64 snapshots were evenly selected, sampling the quenching process from 1,500 K (liquid) to 300 K (room-temperature amorphous phase).
    
    \item \textbf{``iter5'': crystalline-amorphous interface structures equilibrated using GAP.} As the current iteration of the ACE potential could not be used to relax crystalline--amorphous \ce{BaZrS3} interfaces due to unphysical close contacts, a GAP potential was trained on the full ``iter4-3'' dataset. This is in accord with a report that a faster ACE potential requires more training data to reach the same stability as a GAP model trained on a smaller dataset.\cite{bernstein_gap_2024} Two main types of structures were added throughout the 4 separate GAP iterations: 
    \begin{enumerate}
    \item \textit{Manually created crystalline--amorphous interfaces.} Using the NPT protocol used in ``iter4'' with a quench rate of 10$^{13}$ K/s, amorphous structures were generated from three different melt--quench trajectories. An interface was manually created between the amorphous and crystalline material, obtaining three 320-atom unit cells. These structures were relaxed using DFT and added to the dataset. Two of the structures were then equilibrated at 300 K in the NVT ensemble for 5 ps. From each of the two trajectories, 25 evenly-spaced snapshots were selected for labelling. The third structure was equilibrated at 100 K, leading to less thermal motion of the crystalline part of the material, and 5 snapshots from this trajectory were added to the dataset (``iter5-1'').
    \item \textit{Crystalline--amorphous interfaces created by fixing certain regions within the structure in a LAMMPS melt-quench protocol.} The melt-quench protocol described in Ref.~\citenum{zhou_device-scale_2023} was used to fix a region of a 160-atom unit cell while randomising the rest of the cell by melting at 3,000 K. This melting simulation was conducted for 80 ps, based on tests for the mean square displacement (MSD) in LAMMPS, which reached $\approx$ 40 \AA{}$^2$ in each direction ($dx^2$, $dy^2$, $dz^2$). (The same melting time was later used in the melt-quench process to generate the amorphous phase.) Following melting, the temperature was lowered to the experimental annealing temperature of 950 K. The whole structure was then annealed at 950 K in the NVT ensemble, sampling the route to crystallisation. In ``iter5-2'', 100 evenly-spaced snapshots from the melt-quench trajectory, and 125 snapshots from the annealing section were selected and added to the dataset. In ``iter5-3'', the process was repeated without the annealing step, sampling 100 structures with crystalline--amorphous interfaces from the melt--quench protocol. In ``iter5-4'', larger cells were used (360 atoms) and an annealing step was added again, selecting 15 snapshots from the melt-quench trajectory and 40 snapshots from the annealing protocol.
    \end{enumerate}

\end{itemize}

In this study, two versions of the ACE potential were used:
\begin{enumerate}
    \item \textit{ACE ``iter5-4'' trained on low-energy structures} (< 1 eV/atom as calculated by DFT). From the training dataset described in Table \ref{tab:Dataset}, 119 higher-energy structures (17 close-contact dimer structures and 112 RSS structures) were removed. With an energy RMSE of 13.9 meV/atom relative to DFT data, this is a very accurate model that can be used to relax grain boundary structures reasonably well. Although the model can deal with most physically sensible structures, the removal of higher-energy RSS structures and very-close-contact dimers led to the potential failing for the polycrystalline structures generated by Voronoi tesselation (see \hyperref[subsec:GB]{Grain boundary relaxation}). 
    \item \textit{ACE ``iter5-4'' trained on the full dataset, including higher-energy structures}. With an energy RMSE of 23.1 meV/atom, this model fails to relax all grain boundary structures with the same level of accuracy as the lower-energy version of the potential. However, the loss of accuracy is compensated by the flexibility of the potential, which can correctly identify and relax very high energy structures with close contacts between atoms. Given this observation, the relaxation of the polycrystalline cell was performed using this version of the potential.   
\end{enumerate}
We have found that the low-energy version of the ACE model is the best choice for most applications related to the polycrystalline Ba--Zr--S system. However, for the purpose of dealing with high-energy, randomised structures, we recommend the use of the version trained on the full dataset.

\clearpage

\subsection*{Model hyperparameters} \label{subsec:hyp}
\subsubsection*{GAP}
As described in the previous section, the initial training iterations were performed using the Gaussian Approximation Potential (GAP) framework, as implemented in QUIP.\cite{Csanyi2007-py, Kermode2020-wu} The choice is motivated by the previous successful application of the GAP-driven random structure search (GAP-RSS) approach for generating a diverse training dataset and by the GAP model's stability in the low-data regime.\cite{bernstein_novo_2019, bernstein_gap_2024}

To describe atomic environments, we used an expansion into a squared-exponential two-body term and a multibody SOAP term with the relative weighting determined by the specific hyperparameter $\delta$. The chosen contribution of $\delta^{\textnormal{2b}}=1$ and $\delta^{\textnormal{SOAP}}=0.2$ was previously found to be appropriate for the description of short-range interactions (<1 \r{A}) in a complex ternary system.\cite{zhou_device-scale_2023} The width of the Gaussian squared-exponential term was set to 0.5 \AA{}, with a cutoff radius of 5 \AA{}, and we used 15 representative points for the sparsification of the Gaussian process. For the SOAP hyperparameters, the cutoff was set to 5 \AA{}, with a transition width of 0.5 \AA{}, neighbour density smoothness ($\sigma_\textnormal{at}$) of 0.75 \AA{} (as found appropriate for previous work on a GAP-RSS potential for boron \cite{bernstein_novo_2019}), and 2000 representative points. The maximum number of radial and angular basis functions was $l_\textnormal{max}=6$ and $n_\textnormal{max}=10$. The kernel function was chosen as a dot product with exponential $\zeta=4$. The ``expected error'' of the energies, forces, and virials used to label the training data were set by the regularisation parameter, $\sigma$. To prioritise the inclusion of energetically sensible RSS-generated structures, the regularisation parameter is determined based on the energy distance ($\Delta E$) of each structure from the convex hull. Configurations with $\Delta E$ > 20 eV/atom were excluded from the training data. Details regarding each configuration type included in the training data can be found in Table \ref{tab:comparison}.

\subsubsection*{ACE}
Once a sufficiently large dataset was compiled by iteratively training the GAP potential, following iterations could be fit using the \texttt{pacemaker} implementation of ACE.\cite{lysogorskiy_performant_2021} All simulations presented in the main text were performed using the final iteration (``iter5-4'') of the ACE potential, using either the version trained on lower-energy structures or the one based on the full training dataset (see above). A nonlinear Finnis--Sinclair-like embedding with 6 atomic properties and a cutoff of 7.2 \AA{} were found to be necessary for stable simulations of our ternary system. The number of functions per element was set to 1,000. A systematic study of nonlinear functions with different numbers of atomic properties, beyond the scope of this work, could have a further effect on the potential's performance. 

\clearpage

\subsection*{Numerical validation}
\begin{figure}[H]
\centering
  \includegraphics[width=\textwidth]{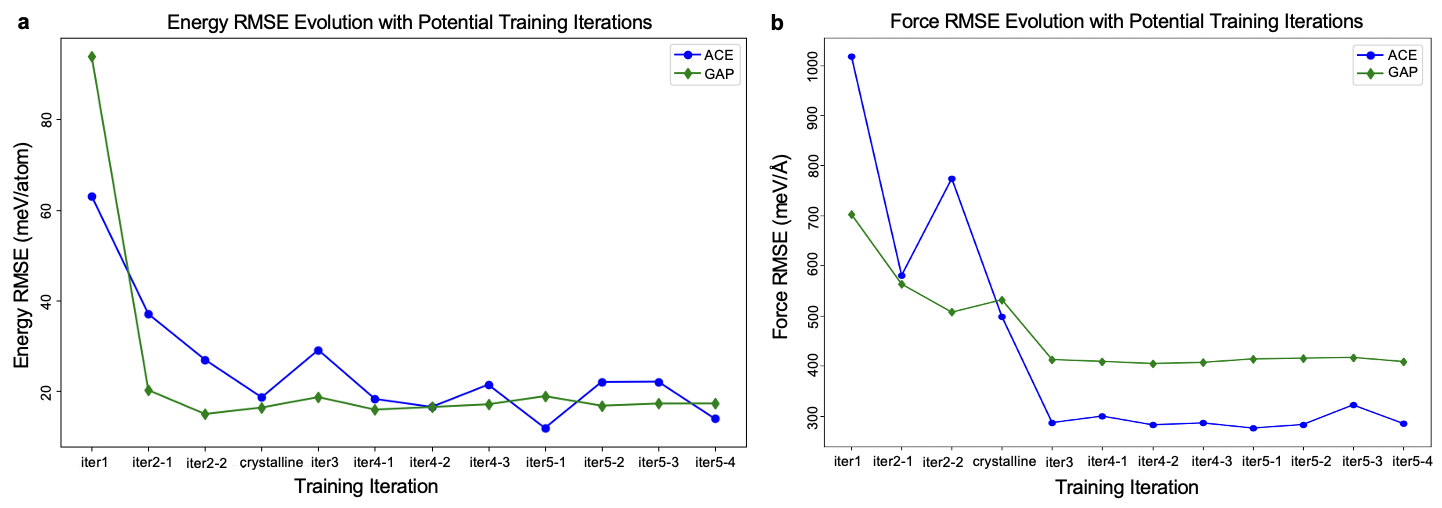}
  \caption{The evolution of (\textbf{a}) the RMSE of the energy and (\textbf{b}) the RMSE of the forces with each training iteration of the GAP and ACE potentials, showing the reduction in error with the addition of training data.}
  \label{fgr:rmse}
\end{figure}

\begin{figure}[H]
\centering
  \includegraphics[width=0.75\textwidth]{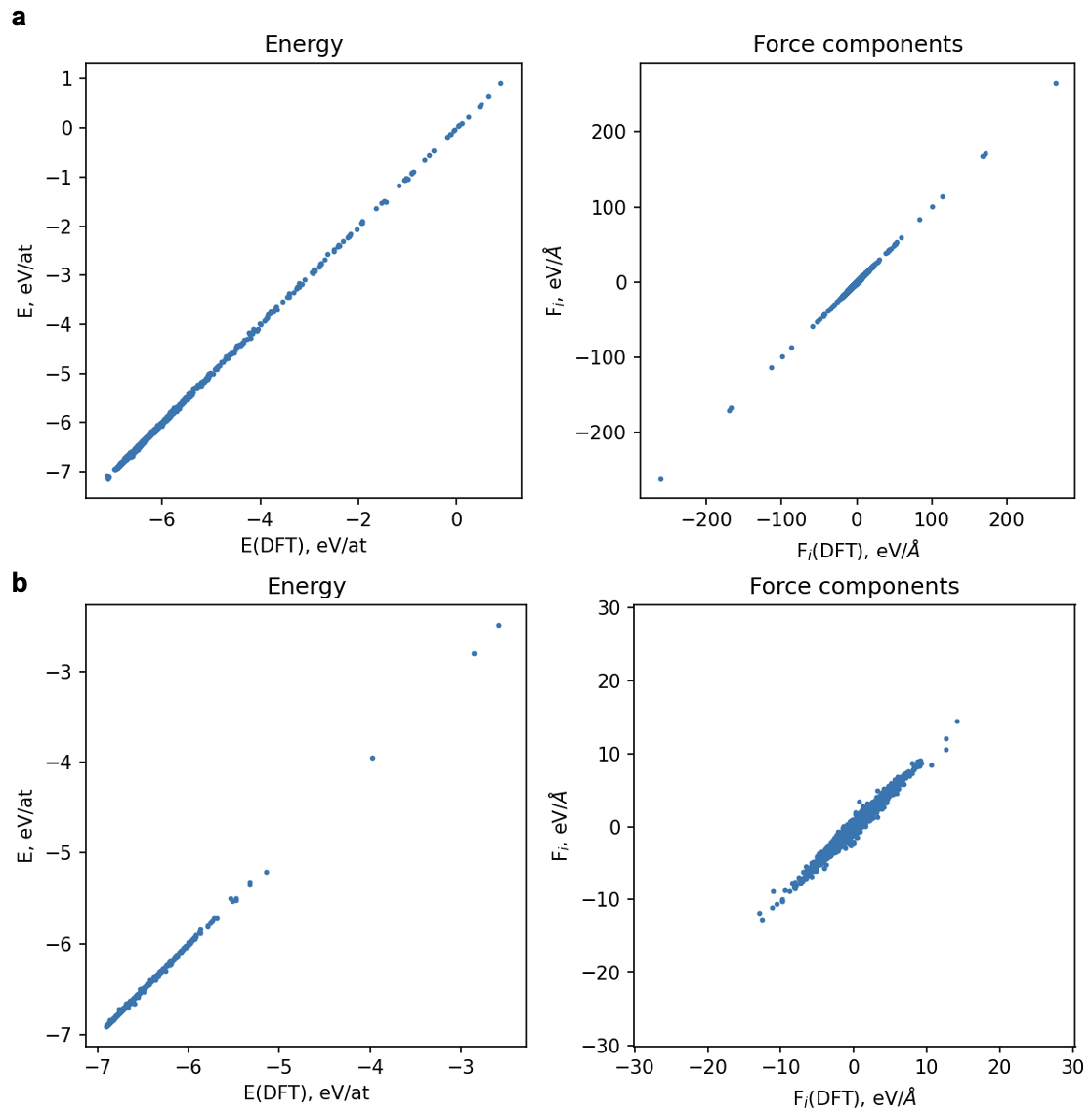}
  \caption{Scatter plots of the energy and forces predicted by DFT single-point energy calculations and by the ACE potential for (\textbf{a}) the training dataset (``iter5-4'') and (\textbf{b}) the test dataset.}
  \label{fgr:pair-plot}
\end{figure}

The performance of each iteration of the GAP and ACE models trained on the same dataset is tested on the final test set, containing 10\% of the structures that were split from the training data during each iteration. Thus, the test set contains all configuration types relevant for the present study, excluding grain boundaries. Overall, the RMSE values between the energies and between the forces calculated using DFT and those obtained using the potential decrease as each configuration type is added to the training set in different iterations (Fig.~\ref{fgr:rmse}). The ACE potential incurs a substantially lower error in forces than the GAP model quite early on in the training (after the addition of the crystalline structures). In the final iteration, the energy RMSE reaches 13.96 meV/atom and that for the forces reaches 284.6 meV/\r{A}. While the ACE potential shows lower RMSE values at the start, it requires a few iterations to become stable in MD simulations. The train and test set scatter plots for the final iteration of the ACE potential are also provided (Fig.~\ref{fgr:pair-plot}). 

\subsection*{DFT computations}
DFT reference data for MLIP training were obtained using the Vienna Ab initio Simulation Package (VASP)\cite{kresse_efficiency_1996, kresse_efficient_1996} with the PBEsol functional. \cite{perdew_restoring_2008} The projector augmented-wave (PAW) pseudopotentials\cite{blochl_projector_1994, kresse_ultrasoft_1999} used corresponded to valence-electron configurations of 5s$^2$5p$^6$6s$^2$ for Ba, 4s$^2$4p$^6$5s$^2$4d$^2$ for Zr and 2s$^2$2p$^4$ for S (PBE, v.5.4). A Monkhorst--Pack grid centred on $\Gamma$ with a $k$-spacing of 0.2 \AA{} was used. The energy cutoff was set to 600 eV with a stopping criterion of 1 $\times$ 10$^{-6}$ eV. Gaussian smearing was applied with a width of 0.05 eV. For the grain-boundary single-point energy computations, the energy tolerance was lowered to 5 $\times$ 10$^{-5}$ eV with $k$-point sampling at $\Gamma$-only to reduce the computational resources required for the large unit cells.

\subsection*{Molecular dynamics melt-quench protocol} \label{subsec:mq-protocol}
A timestep of 1 fs and a Nos\'e{}--Hoover thermostat were used for all MD simulations, which were conducted using LAMMPS.
To obtain the amorphous phase starting from the crystalline structure, a melt--quench MD simulation was performed in the NPT ensemble, allowing the structure to first melt at 3,000 K (holding for 80 ps), then quenching to 1,500 K at a rate of 10$^{14}$ K/s, equilibrating for 80 ps and performing a quench to 300 K at a rate of 10$^{13}$ K/s. The structure was finally annealed at 300 K for 80 ps. 

The effect of the quench rate for the 1,500 K $\rightarrow$ 300 K part of the protocol on the resulting density and coordination numbers is analysed in Table S3. 

\subsection*{Grain boundary relaxation} \label{subsec:GB}
The multi-grain structures were created using Atomsk \cite{hirel_atomsk_2015}, which constructs polycrystalline models with random grain orientations using the Voronoi tesselation method. 
For the 2D grain boundary systems, models of coincidence site lattice GBs for an orthorhombic system with \textit{a}:\textit{b} ratio of 0.98 for rotations around [001] were constructed according to the parameters in Table \ref{tab:Grain boundaries}, based on Ref.~\citenum{king_geometrical_1993}. The unit cells were chosen to be larger than 40 \r{A} $\times$ 40 \r{A} in the directions perpendicular to the grain boundary to avoid interactions due to periodic boundary conditions. Overlapping atoms at a distance closer than 1.5 \r{A} were deleted from the generated structures. The grain boundary energies were minimised using an energy tolerance of 10$^{-6}$ eV (10$^{-7}$ eV for the polycrystalline model) using the conjugate gradient minimisation algorithm implemented in LAMMPS. The polycrystalline structures were then further annealed at 300 K for 200 ps and quenched down to 0.1 K across 300 ps, following the protocol established by Van Swygenhoven et al. (Ref.~\citenum{van_swygenhoven_grain-boundary_2000}). Similar settings and a similar protocol were used to relax the bulk crystalline structure.

\subsection*{X-ray diffraction pattern simulations}
The XRD pattern simulations were generated using the Debye Calculator Python package.\cite{Johansen2024} To convert between \textit{Q} and $2\theta$, the use of Cu K-$\alpha$ radiation with a wavelength of 1.5406 \r{A} was assumed. An isotropic atomic displacement (Debye--Waller) factor of 0.3 \r{A}$^2$ was used.

\subsection*{Visualisation}
The similarity maps for the training dataset (see Fig.~1b in the main text) were constructed using the SOAP similarity kernel to quantify the similarity between structures. Structural similarity was determined by averaging the atomic environment descriptor vectors across each cell. Dimensionality reduction was performed using the Uniform Manifold Approximation and Projection (UMAP) algorithm, with a minimum distance (\texttt{min\_dist}) of 0.5 between points in the reduced space and a maximum of 10 nearest neighbors (\texttt{n\_neighbors}).

All structural images were produced using OVITO.\cite{stukowski_visualization_2009} The OVITO Python package was additionally used to obtain the average value of the bond lengths and coordination numbers of the amorphous phase, assuming a cutoff radius of 3.1 \r{A} in the case of Zr--S and 3.8 \r{A} in the case of Ba--S.

\clearpage

\section*{Supplementary tables}

\begin{table}[!h]
\centering
\caption{Composition of the training dataset.}
\begin{tabular}{llccc}
\toprule
\multicolumn{2}{l}{} & \multicolumn{2}{c}{Dataset size} \\
\cmidrule(rl){3-4}
 {Training data} &  & {Cells} & {Atoms} \\
\midrule
Free atoms & Ba/Zr/S & 3 & 3 \\
\hline
Dimers & Ba--Ba/Zr--Zr/S--S/Ba--S/Ba--Zr/Zr--S & 72 & 144 \\
\hline
iter0 & AIRSS structures & 90 & 2025 \\
\hline
\multirow{10}{*}{\shortstack{iter1-1 \\ iter1-2 \\ iter1-3 \\ iter1-4 \\ iter1-5\\ iter1-6\\ iter1-7\\iter1-8\\iter1-9\\iter1-10}} &  &  & \\
 &  &  &   \\
 &  &  &    \\
 &  &  &   \\
 & GAP-driven RSS & 899 & 7305 \\
&  &  &   \\
&  &  &   \\
&  &  &   \\
&  &  &  \\
&  &  &   \\
\hline
iter2-1 \& iter2-2 & NVT MD (from RSS) & 180 & 18000  \\
\hline
+ & Crystalline structures & 97 & 15520  \\
\hline
iter3 & NVT melting (from crystalline) & 405 & 64800 \\
\hline
iter4-1 \& iter4-2 \& iter4-3 & NPT trajectories using ACE & 417 & 78120 \\
\hline
\multirow{5}{*}{\shortstack{iter5-1 \\ iter5-2 \\ iter5-3 \\ iter5-4}} &  &  &  \\
&  &  &  &  \\
& Annealed cryst.--amorphous interfaces & 618 & 116703 \\
&  &  &  \\
&  &  &  \\
\hline
\textbf{Total} & & \textbf{2781} & \textbf{302620} & \\
\bottomrule
\label{tab:Dataset}
\end{tabular}
\end{table}

\begin{table}[h!] 
\centering
\small
\caption{Regularisation parameters (``expected errors'') for each configuration type in the GAP training dataset.}
\begin{tabular}{lccc}
\toprule
\textbf{Configuration type} & $\mathbf{\sigma_E}$ & $\mathbf{\sigma_F}$ & $\mathbf{\sigma_v}$ \\ \hline
Isolated atoms (\AA) & 0.0001 & - & - \\ \hline
Dimers & 0.1 & 0.5 & - \\ \hline
RSS & \makecell[l]{
$dE \leq 0.2$: $\sigma_E = 0.001$ \\ 
$0.2 < dE \leq 1$: linear ramp (0.001 to 0.1) \\ 
$dE > 1$: $\sigma_E = 0.1$
} & \makecell[l]{
$\sigma_F = \sqrt{\sigma_E}$
} & \makecell[l]{
$\sigma_v = 2.0 \times \sqrt{\sigma_E}$
} \\ \hline
\makecell[l]{Domain-specific \\
(crystalline structures, \\ NPT, NVT etc)} & 0.001 & 0.02 & 0.05 \\ \bottomrule
\end{tabular}
\label{tab:comparison}
\end{table}

\begin{table}[htbp]
\centering
\small
\setlength{\tabcolsep}{6pt}
\renewcommand{\arraystretch}{1.2}
\caption{Summary of densities and coordination statistics for structural models of amorphous \ce{BaZrS3} obtained from melt--quench MD simulations (see details in the corresponding part of the ``Computational methods'' section, p.\ S6). We report results for different quench rates for quenching from 1,500 to 300~K; the results for the $10^{13}$ K/s quench rate in this part of the simulation are shown in the main text.}
\begin{tabular}{l c c c c}
\toprule
\textbf{Quench rate} & $10^{12}$ K/s & $10^{13}$ K/s & $10^{14}$ K/s & $10^{15}$ K/s  \\
\hline
\textbf{Density (g/cm\textsuperscript{3})} & 3.96 & 3.94 & 3.90 & 3.63 \\
\hline
\textbf{Average Zr CN} & 5.91 & 5.88 & 5.83 & 5.15 \\
\hline
\multicolumn{5}{c}{\textbf{Zr CN distribution (\% of Zr atoms)}} \\
\hline
CN = 3 & - & - & - & 0.1 \\
CN = 4 & 0.3 & 0.2 & 0.4 & 7.0 \\
CN = 5 & 13.7 & 16.2 & 20.1 & 46.1 \\
CN = 6 & 80.5 & 77.1 & 74.9 & 45.4 \\
CN = 7 & 5.5 & 6.3 & 4.4 & 1.4 \\
CN = 8 & - & 0.2 & 0.2 & - \\
\hline
\multicolumn{5}{c}{\textbf{Ba CN distribution (\% of Ba atoms)}} \\
\hline
CN = 3 & - & - & - & 0.7 \\
CN = 4 & 0.1 & 0.1 & 0.1 & 4.0 \\
CN = 5 & 1.1 & 1.1 & 2.4 & 19.5 \\
CN = 6 & 13.0 & 14.8 & 17.2 & 35.7 \\
CN = 7 & 36.0 & 39.8 & 41.2 & 29.3 \\
CN = 8 & 38.3 & 33.9 & 31.4 & 10.2 \\
CN = 9 & 10.5 & 9.9 & 7.2 & 0.6 \\
CN = 10 & 1.0 & 0.4 & 0.5 & - \\
\bottomrule
\end{tabular}
\label{tab:coordination_statistics}
\end{table}

\begin{table}[!h]
\caption{Comparison of simulation and experimental (EXAFS) values of the bond length and coordination number of Zr in the amorphous and polycrystalline phases (expt.\ data from Ref.~\citenum{mukherjee_interplay_2023}). Voronoi tesselation was used to create a polycrystalline model with 6 randomly placed grains in a 250 $\times$ 250 $\times$ 250 supercell (615,214 atoms). The cut-off used for calculating the average CN and bond length between Zr and S atoms was 3.1 \r{A}.}
\centering
\begin{tabular}{l}
\bottomrule
\textbf{Amorphous \ce{BaZrS3}} \\ 
\hline
$d$(Zr--S) (\AA): \\ 
\hspace{1cm} Simulation: 2.575 \\ 
\hspace{1cm} Experiment: 2.593 $\pm$ 0.006 \\ 
\hline
Average CN(Zr): \\ 
\hspace{1cm} Simulation: 5.88 \\ 
\hspace{1cm} Experiment: 5.2 $\pm$ 0.04 \\ 
\hline
\textbf{Polycrystalline \ce{BaZrS3}} \\ 
\hline
$d$(Zr--S) (\AA): \\ 
\hspace{1cm} Simulation: 2.540 \\ 
\hspace{1cm} Experiment: 2.545 $\pm$ 0.003 \\ 
\hline
Average CN(Zr): \\ 
\hspace{1cm} Simulation: 5.87 \\ 
\hspace{1cm} Experiment: 6 $\pm$ 0.2 \\ 
\bottomrule
\end{tabular}
\label{tab:comparison-values-poly}
\end{table}

\begin{table}[!h]
\caption{Comparison between the experimentally-determined lattice parameters of crystalline \ce{BaZrS3} extrapolated to 0 K \cite{jaykhedkar_how_2023} and the lattice parameters of the structure relaxed using the ACE potential and the Broyden–Fletcher–Goldfarb–Shanno (BFGS) algorithm with a line search mechanism implemented in the ASE Python package.}
\centering
\begin{tabular}{lccc}
\toprule
 & \textbf{\textit{a} (\AA{})} & \textbf{\textit{b} (\AA{})} & \textbf{\textit{c} (\AA{})} \\ \hline
Experiment & 7.075 &  6.962 & 9.941 \\ 
\hline
Simulation & 7.057 & 7.163 & 10.050 \\ 
\bottomrule
\end{tabular}
\label{tab:Grain boundaries}
\end{table}

\begin{table}[!h]
\caption{Expected stable grain boundaries of the relaxed \ce{BaZrS3} structure based on the coincidence site lattice model for an orthorhombic unit cell according to Ref.~\citenum{king_geometrical_1993}.}
\centering
\begin{tabular}{ccl}
\toprule
\textbf{$\Sigma$} & \textbf{$\theta$} ($\degree$) & \textbf{Number of atoms}  \\ \hline
$\Sigma$25 & 78.463 & 796 atoms\\ \hline
$\Sigma$31 & 52.200 & 1224 atoms \\ \hline
$\Sigma$35 & 44.415 & 1102 atoms \\ 
\bottomrule
\end{tabular}
\label{tab:Grain boundaries}
\end{table}

\clearpage

\section*{Supplementary figures}

\begin{figure}[H]
\centering
  \includegraphics[width=\textwidth]{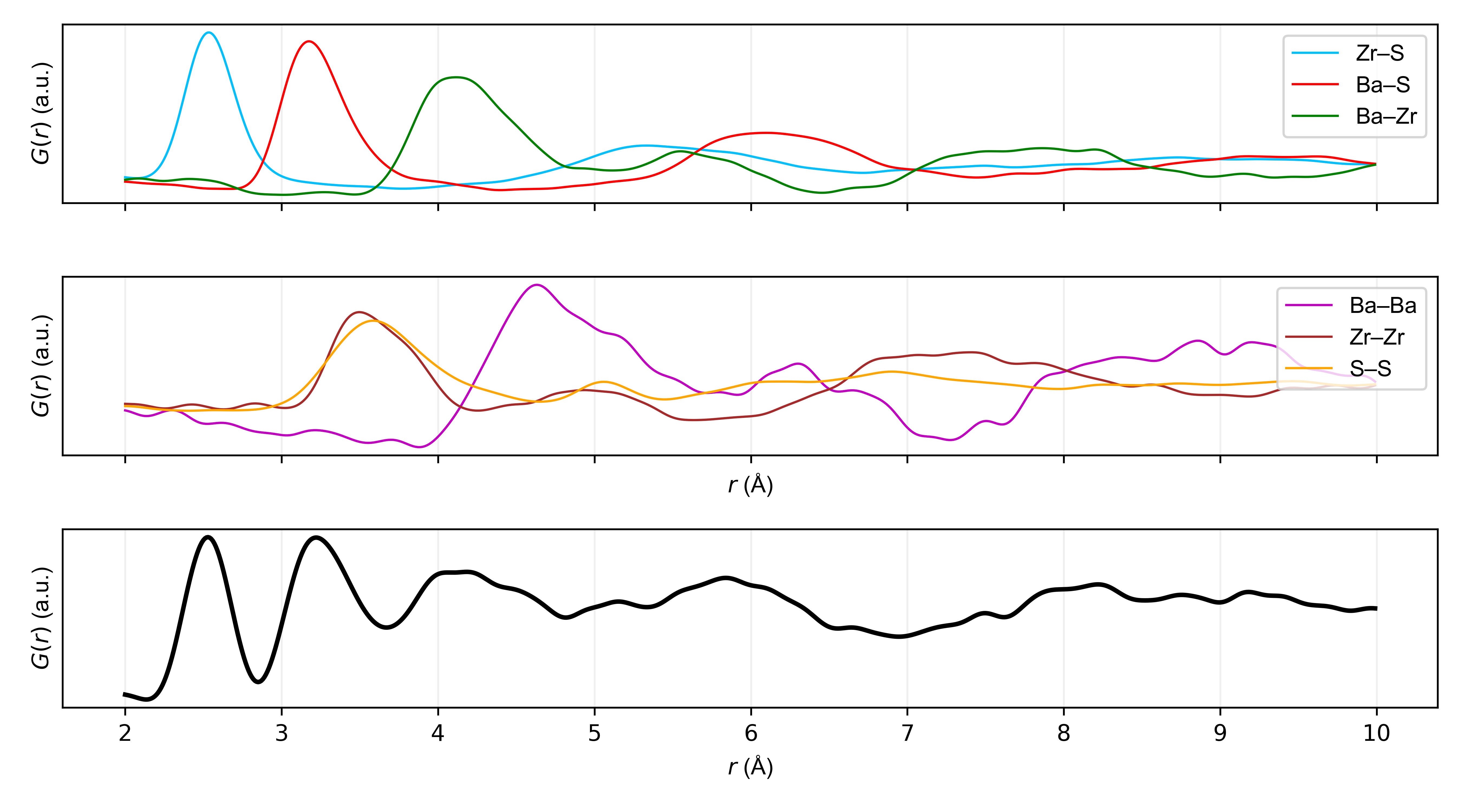}
  \caption{Partial and total pair distribution functions for the simulated amorphous structure, extended to 20 \r{A}.}
  \label{fgr:pdf}
\end{figure}

\begin{figure}[H]
\centering
  \includegraphics[width=0.5\textwidth]{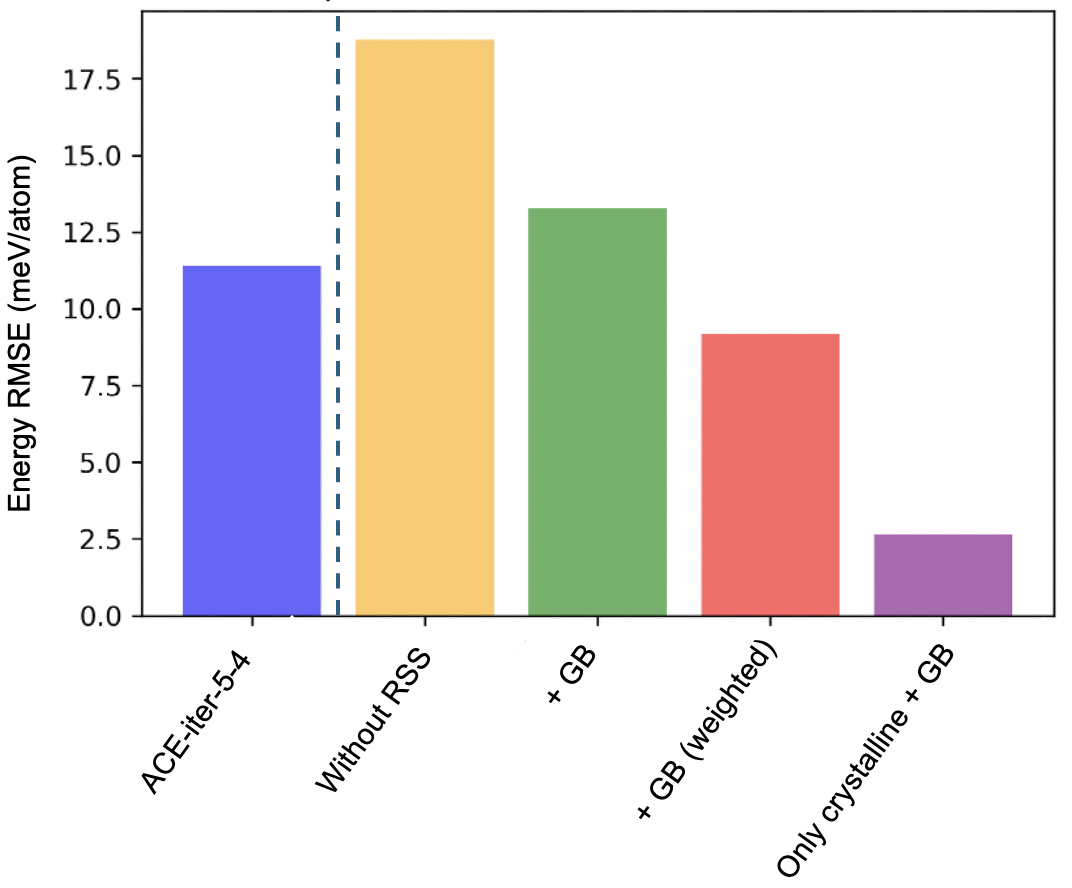}
  \caption{The effect of modifying the iter5-4 training dataset (containing only structures with energies below 1 eV/atom) on the RMSE of the relaxed grain boundaries calculated by DFT. The inclusion of RSS structures leads to a lower RMSE, while the simple addition of the grain-boundary structures to the training dataset does not improve the RMSE without increasing the weighting of the GB and crystalline structures. The RMSE is lowest for a potential trained only on crystalline and GB structures.}
  \label{fgr:ablation}
\end{figure}

\begin{figure}[H]
\centering
  \includegraphics[width=0.85\textwidth]{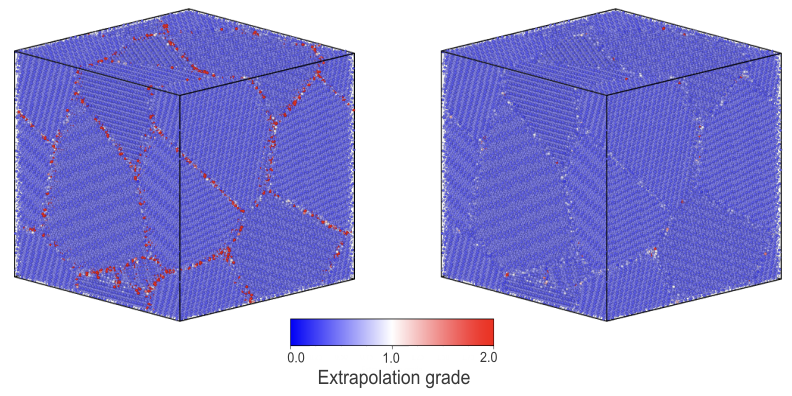}
  \caption{Extrapolation grade (based on the D-optimality/MaxVol algorithm \cite{lysogorskiy_active_2023}) calculated for each atom in the unrelaxed and relaxed polycrystalline model with 6 randomly-oriented grains (613,229 atoms). A value between 0 and 1 means that the potential is in the ``interpolation region'', while a value between 1 and 2 places it in the ``accurate extrapolation region''.\cite{ito_machine_2024} The unrelaxed structure presents many atoms with grades of 2 and above in the boundary region, whereas the relaxed structure does not contain atoms with grades above the accurate extrapolation region.}
  \label{fgr:extrapolation-grade}
\end{figure}

\begin{figure}[H]
\centering
  \includegraphics[width=\textwidth]{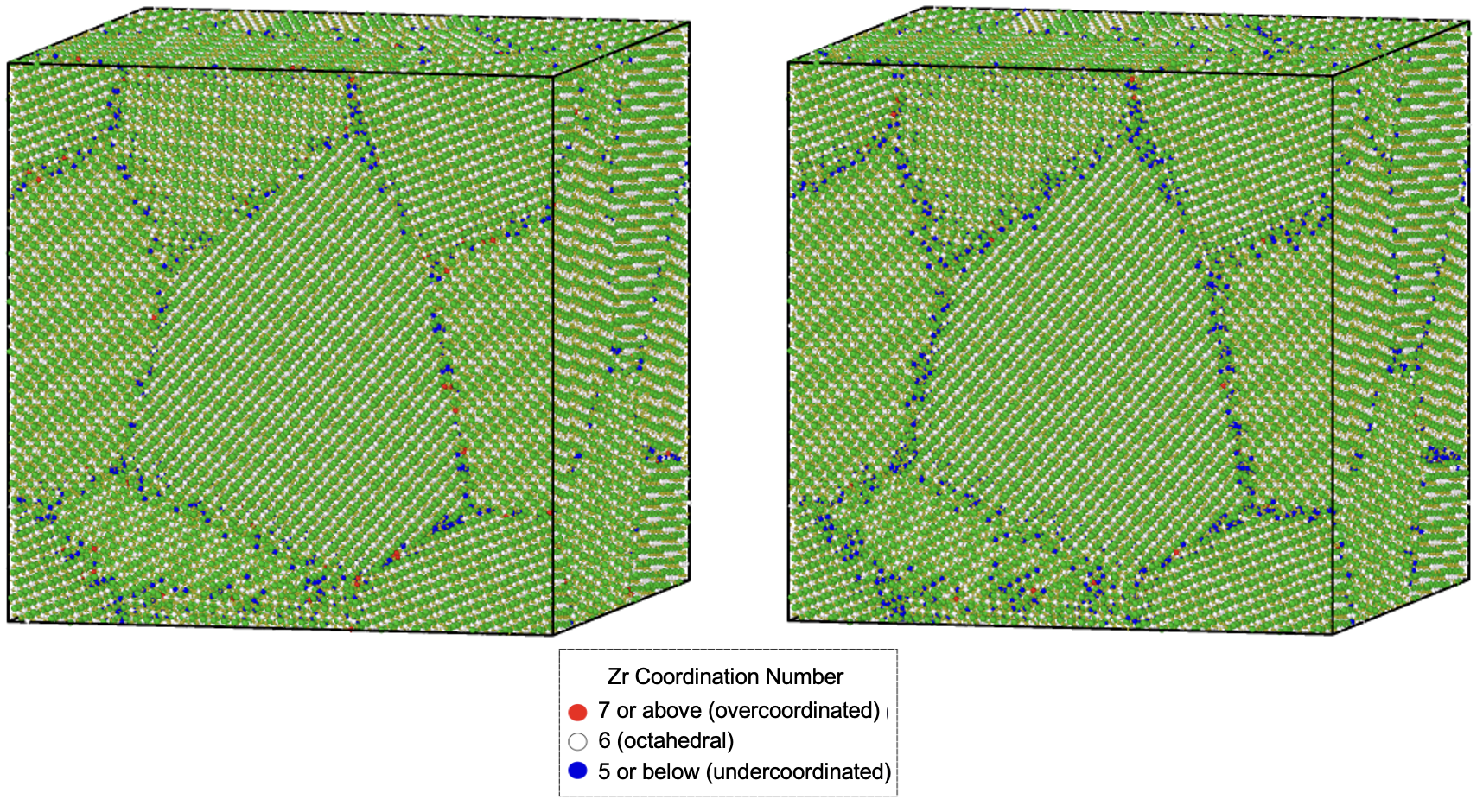}
  \caption{The coordination around the Zr atom in the unrelaxed and the relaxed polycrystalline \ce{BaZrS3} structure with 6 randomly-oriented grains (613,229 atoms).  }
  \label{fgr:coord-crystalline}
\end{figure}

\begin{figure}[H]
\centering
  \includegraphics[width=0.75\linewidth]{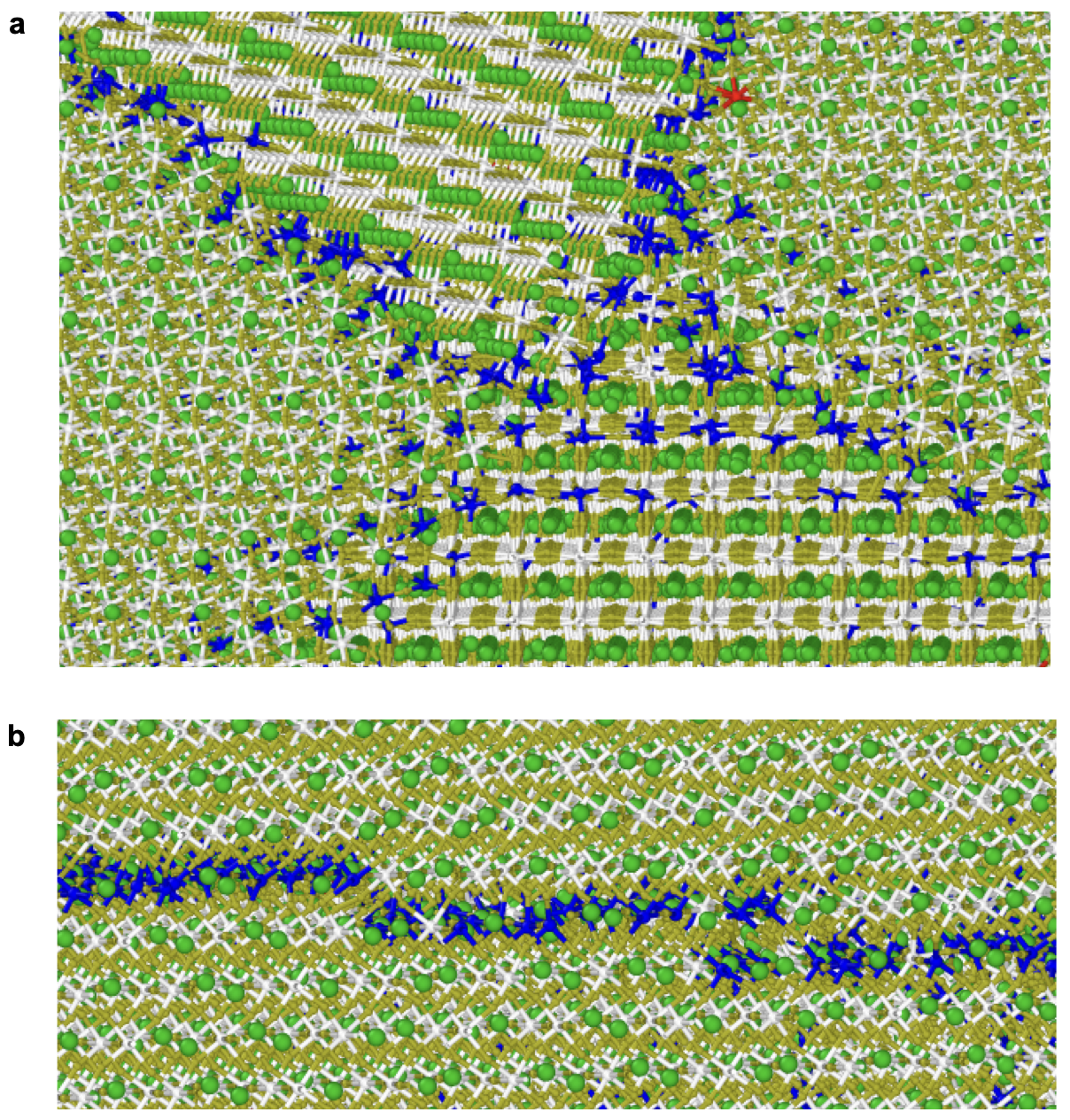}
  \caption{Detail of atomistic structure at the grain boundaries in a polycrystalline model (6 grains, 615,129 atoms). (\textbf{a}) Intersection of four randomly-oriented grains. (\textbf{b}) Dislocation along the grain boundary. Undercoordinated (5-coordinate or below) Zr atoms are shown in blue, octahedral (6-coordinate) in white, and overcoordinated (7-coordinate or above) in red.}
  \label{fgr:xrd-polycrystal-single-crystal}
\end{figure}

\begin{figure}[H]
\centering
  \includegraphics[width=\textwidth]{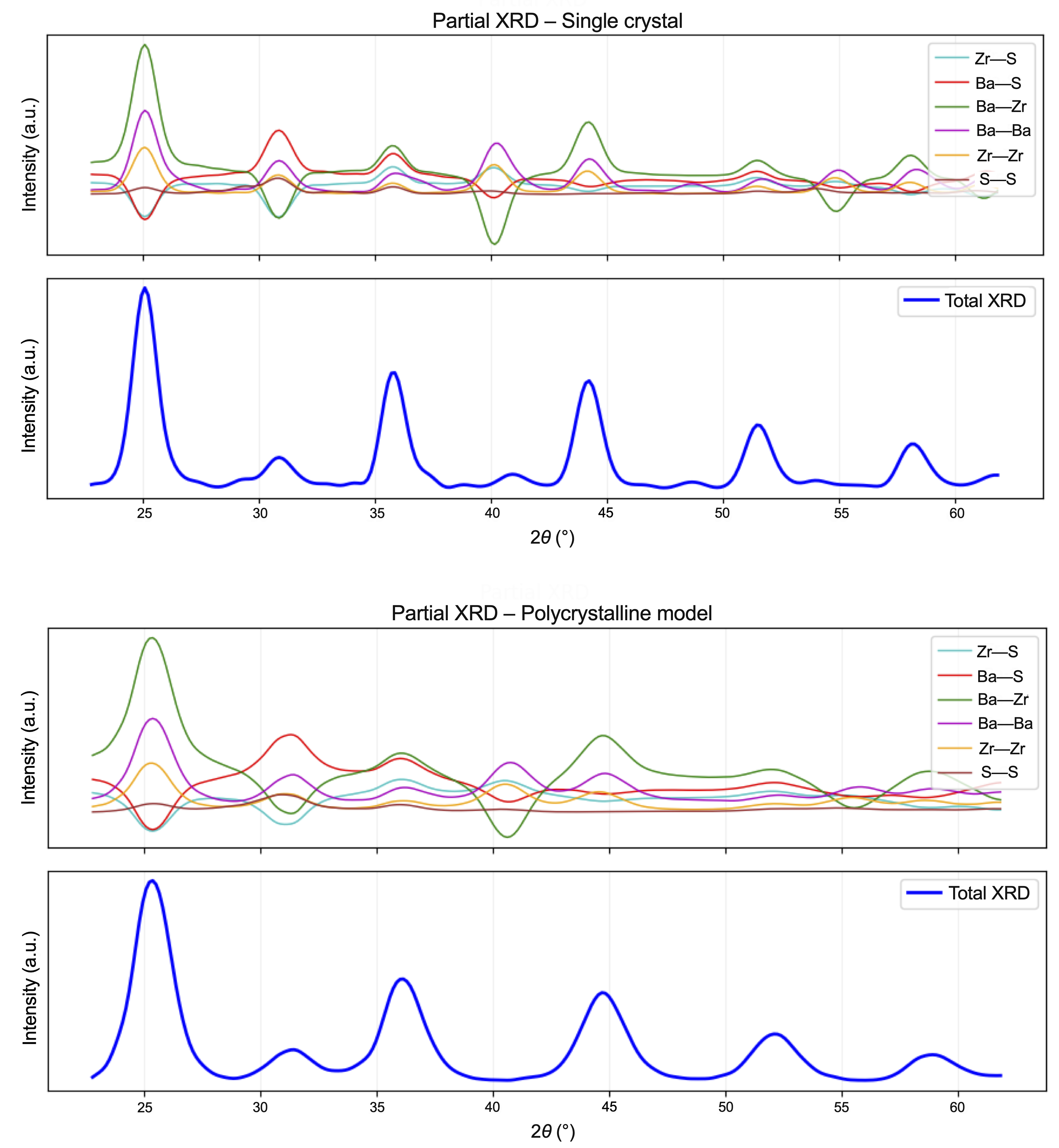}
  \caption{Comparison of the partial and total XRD patterns of a single-crystal (10,240-atom) and polycrystalline model of \ce{BaZrS3} containing 8 randomly-oriented grains in a box with a side length of 10 nm.}
  \label{fgr:partial-XRD}
\end{figure}

\clearpage

\section*{Supplementary references}
\vspace{3mm}

\providecommand*{\mcitethebibliography}{\thebibliography}
\csname @ifundefined\endcsname{endmcitethebibliography}
{\let\endmcitethebibliography\endthebibliography}{}